\documentstyle[preprint,aps,prb]{revtex} 
\begin{document}
\draft
\preprint{\today}

\title{First-principles theory of ferroelectric phase transitions for
perovskites:\\ The case of BaTiO$_3$}

\author{W.~Zhong and David Vanderbilt}

\address{Department of Physics and Astronomy,
  Rutgers University, Piscataway, NJ 08855-0849}

\author{K.~M.~Rabe}

\address{Department of Applied Physics, Yale University, New Haven, CT
06520-8284}
\date{}
\maketitle

\begin{abstract}
We carry out a completely first-principles study of the ferroelectric
phase transitions in BaTiO$_3$. Our approach takes advantage of two
features of these transitions: the structural changes are small,
and only low-energy distortions are important.  Based on these
observations, we make systematically improvable approximations
which enable the parameterization of the complicated energy
surface.  The parameters are determined from first-principles
total-energy calculations using ultra-soft pseudopotentials and a
preconditioned conjugate-gradient scheme.  The resulting effective
Hamiltonian is then solved by Monte Carlo simulation.  The
calculated phase sequence, transition temperatures, latent heats,
and spontaneous polarizations are all in good agreement with
experiment.  We find the transitions to be intermediate between
order-disorder and displacive character.  We find all three phase
transitions to be of first order.  The roles of different
interactions are discussed.
\end{abstract}


\narrowtext

\section{Introduction}
Because of their simple crystal structure, the perovskite oxides present a
special opportunity for the development of a
detailed theoretical understanding of the ferroelectric phase
transition.  Within this family of materials, one finds transitions
to a wide variety of low-symmetry phases, including ferroelectric
and antiferroelectric transitions. Both first- and second-order
transitions are observed, with a full spectrum of transition
behavior ranging from displacive to order-disorder.  The properties
of BaTiO$_3$, a much-studied prototypical example of this class of
compounds,\cite{lines} exemplify this rich behavior.  BaTiO$_3$
undergoes a succession of first-order phase transitions, from the
high-temperature high-symmetry cubic perovskite phase
(Fig.\ \ref{stru}) to slightly distorted ferroelectric structures
with tetragonal, orthorhombic and rhombohedral symmetry.  There is
increasing evidence that the cubic-to-tetragonal transition, at first
thought to be of the simple displacive kind, may instead be better
described as of the order-disorder type.

The variety exhibited by the perovskite oxides shows that the phase
transformation behavior depends on details of the chemistry and
structural energetics of each particular compound.  Therefore, it
is of the first importance to develop a microscopic theory of the
materials properties which determine the ordering of the phases,
the character and thermodynamic order of the transitions, and the
transition temperatures.  The value of a microscopic approach has
long been appreciated, but its realization was hindered by the
difficulty of determining microscopic parameters for individual
compounds.  The forms of phenomenological model Hamiltonians
\cite{lines,dove,pytt,cowl} were limited by the available
experimental data, leading to oversimplification and ambiguities in
interpretation.  For the perovskite oxides, empirical \cite{bilz}
and non-empirical pair potential methods \cite{boyer} did not offer
the high accuracy needed for the construction of realistic models.

First-principles density functional calculations offer an
attractive approach for enhancing our microscopic understanding of
perovskites and other ferroelectrics.  The all-electron
Full-potential Linearized-Augmented-Plane-Wave (FLAPW) method has
been used by several groups to study ferroelectricity in perovskites
within the Local Density Approximation (LDA).\cite{cohen,singh}
Recently, King-Smith and Vanderbilt performed a systematic
study of structural and dynamical properties and energy surfaces for
eight common perovskites,\cite{king1,king11} using the first-principles
ultra-soft pseudopotential method and the LDA. These calculations
demonstrate that ferroelectricity in the perovskites reflects a
delicate balance between long-range electrostatic forces which
favor the ferroelectric state, and short-range repulsions which
favor the cubic phase. While constrained to calculations of
zero-temperature properties, these calculations yield correct
predictions of ground state structures and occurrence of
ferroelectric phases for certain materials.  They show that
high-quality LDA calculations can provide considerable insight into
the nature of the total-energy surface in the perovskites.  For
further insight into the energetics of ferroelectric compounds, the
polarization generated by various distortions can be studied
directly, using a recent first-principles method by King-Smith and
Vanderbilt.\cite{king2}  This approach has been applied to the
investigation of the zone-center phonons in the common perovskite
oxides.\cite{zhong}

The application of these first-principles methods can clearly form
a foundation for the realistic study of the finite-temperature
phase transitions.  While an {\it ab-initio} molecular-dynamics
simulation of the structural phase transition is not
computationally feasible at present, we purse an alternative
first-principles approach to study ferroelectric phase transitions
and demonstrate its application to BaTiO$_3$.  In particular, we
(i) construct an effective Hamiltonian to describe the important
degrees of freedom of the system,\cite{rabe,rabeuw} (ii) determine
all the parameters of this effective Hamiltonian from high-accuracy
{\it ab-initio} LDA calculations,\cite{king1,zhong,vand1} and
(iii) carry out Monte Carlo (MC) simulations to determine the phase
transformation behavior of the resulting system.
An abbreviated presentation of this work has already
appeared in Ref.\ \onlinecite{whenshallwethreemeetagain}.

The remainder of this paper is organized as follows.  In Sec.~II,
we go through the detailed procedure for the construction of the
effective Hamiltonian, and give the explicit formula.  In Sec.~III,
we describe our first-principles calculations and the determination
of the expansion parameters in the Hamiltonian.  The technical
details of the Monte Carlo simulation are presented in Sec.~IV.  In
Sec. V, we report our calculated transition temperatures, order
parameters, and phase diagram, as well as thermodynamic order and
nature of the phase transitions.  The role of different interactions
in determining the phase transition behavior is also discussed.
Sec.~VI concludes the paper.

\section{Construction of the Hamiltonian}

\subsection{Approximations and local modes}
\label{sec:approx}

The central quantity for studying the equilibrium properties of a
system at finite temperature is its partition function.  This can be
determined from the energy surface, i.e., the total potential
energy as a functional of the atomic coordinates.  Since the
contribution to the partition function decreases exponentially with
increasing energy, it is possible to obtain an accurate partition
function for low-temperature applications from a simplified energy
surface including only low-energy configurations.  Our goal is to
construct a parameterized Hamiltonian which (i) is {\it ab initio},
involving no empirical or semi-empirical input; (ii) results in an
accurate partition function for the temperature range of interest;
(iii) is fully specified by a few {\it ab-initio} total energy
calculations; and (iv) involves only approximations that are
systematically improvable and removable.

Our first fundamental approximation is to use an energy surface
represented by a low-order Taylor expansion.  Both experiments and
first-principles total energy calculations suggest that the
ferroelectric (FE) phase transition involves only very small atomic
displacements and strain deformations from the equilibrium cubic
structure. It is reasonable to assume that all the atomic
configurations with significant contribution to the partition function
would be close to this cubic structure in the temperature range of
interest.  Thus, it is natural to represent the energy surface by a
Taylor series in the displacements from the cubic structure.  We
include up to fourth-order terms in our expansion; this is clearly a
minimum, since ferroelectricity is intrinsically an anharmonic
phenomenon.  By including higher-order terms, this approximation could
later be systematically improved.

It is convenient to describe the small distortions from the cubic
structure in terms of the three acoustic and twelve optical
normal-mode coordinates per k-point.  While this could be
regarded as only a change of basis, it motivates our second
fundamental approximation, which is to restrict the expansion to
include only low-energy distortions.  To achieve this separation,
we note that both experimentally-measured and LDA-calculated
phonon dispersion relations\cite{lines,zhong} show that only
the lowest TO modes (soft modes) and long-wavelength acoustic
phonons (strain variables) make
significant contributions to the phonon density of states at low
energy.  Experimental studies also suggest that the FE phase
transitions are accompanied by a softening of the lowest TO mode
and the appearance of a strain.
All other phonons are hardly affected by the
transitions. It is then our second approximation to
express the energy surface only as a function of the soft-mode
amplitudes and strain.
This approximation reduces the number of degrees of freedom
per cell from fifteen to six, and greatly reduces the number of
interaction parameters needed.  If necessary, this approximation
could later be relaxed by including additional modes.

It is convenient to describe the soft mode over
the whole Brillouin zone (BZ) in terms of a collective
motion of ``local modes,'' just as one describes an acoustic phonon
in terms of a collective displacement of individual atoms.
However, there is more than one choice of local mode which will
generate the same soft mode throughout the BZ; an intelligent choice
can simplify the Hamiltonian and reduce the number of calculations
needed.\cite{rabepapers}  First, the local mode should be as
symmetric as possible, so as to minimize the number of expansion
parameters needed.  Second, the
interactions between local modes at different sites are more
difficult to treat than their on-site energy, so the local
mode should be chosen so as to minimize intersite interactions.
%
For perovskite ABO$_3$ compounds, the highest symmetry is achieved by
centering the local mode on either atom A or B.  In the case
of BaTiO$_3$, the Ti--O bond is much stronger than the Ba--O bond
and the motion of the Ti is more important in the FE transition, so
we choose the local mode which is centered on the Ti atom.

The soft zone-center ($\bf k$=0) FE mode in BaTiO$_3$ is a
$\Gamma_{15}$ mode which can be characterized by the four parameters
$\xi_A$, $\xi_B$, $\xi_{{\rm O}\parallel}$, and $\xi_{{\rm O}\perp}$
(for a mode polarized along the $j$'th Cartesian direction, these refer
to the displacements of the A atom, the B atom, the O atom that
forms a B--O bond along direction $j$, and the other two O atoms,
respectively).  We take the corresponding local mode to consist of a
motion of the central A atom by amount $\xi_A$, the eight
neighboring B atoms by amounts $\xi_B/8$, and the six neighboring O
atoms by amounts $\xi_{{\rm O}\parallel}/2$ or $\xi_{{\rm O}\perp}/2$,
along the $j$'th Cartesian direction.  This mode is
illustrated in Fig.~(\ref{stru}); its amplitude is denoted $u_j$.
An arbitrary $k=0$ soft mode can then be realized as a linear
superposition of these local modes having identical amplitudes
$(u_x,u_y,u_z)=\bf u$ in every cell.

The harmonic interactions between the local-mode amplitudes ${\bf
u}_i$ connecting neighboring cells $i$ must be chosen to reproduce
the harmonic behavior of the soft-mode branch throughout the BZ.
Long-range Coulomb forces are known to play an important part in
these interactions; they are characterized in terms of the
calculated Born (or ``transverse'') effective charges.\cite{zhong}
Thus, the harmonic intersite interactions are represented by a sum
of two contributions: an infinite-range piece that is precisely the
interaction of point dipoles whose magnitude is given by the Born
effective charge, and corrections which we take to be of covalent
origin and therefore local.

To be completely general, anharmonic intercell interactions between
neighboring ${\bf u}_i$ would likewise have to be included.  Instead,
we include only {\it on-site} anharmonic interactions, which are chosen
in such a way that the anharmonic couplings for $k=0$ modes of the real
system are correctly reproduced.  This ``local anharmonicity
approximation'' is an important feature which helps make our scheme
tractable and efficient.  To go beyond this approximation, one could
carry out a careful series of frozen-phonon LDA calculations on
supercells to determine anharmonic couplings at other points in the
BZ.  However, past experience has shown that calculations of this kind
are very cumbersome because of the large number of parameters which
have to be determined.\cite{narasimhan}

With these approximations, our Hamiltonian
consists of five parts: a local-mode self energy, a long-range
dipole-dipole interaction, a short-range interaction
between soft modes, an elastic energy, and an interaction between the
local modes and local strain.  Symbolically,
\begin{eqnarray}
E^{\rm tot} & = & E^{\rm self}(\{ {\bf u} \} )+ E^{\rm dpl}
(\{ {\bf u}  \} )+ E^{\rm short} (\{ {\bf u}  \} ) \nonumber\\
& + & E^{\rm elas} (\{ \eta_l \}) +E^{\rm int} (\{ {\bf u}\},
\{\eta_l \})\; ,
\end{eqnarray}
where ${\bf u}$ is the local soft-mode amplitude vector, and $\eta_l$
is the six-component local strain tensor in Voigt notation ($\eta_1 =
e_{11}$, $\eta_4 = 2 e_{23}$).   In the following
subsections, we present the explicit formulae for these five
contributions.

\subsection{Local mode self energy}

The first term is
\begin{equation}
E^{\rm self}(\{ {\bf u} \}) = \sum_i E({\bf u}_i) \;,
\label{eqself}
\end{equation}
where $E({\bf u}_i)$ is the energy of an isolated local mode at cell
${\bf R}_i$ with amplitude ${\bf u}_i$, relative to that of the perfect
cubic structure.  To describe the FE phase, $E({\bf u}_i)$ must contain
anharmonic as well as harmonic contributions.  Since the reference
structure is cubic, only even-order terms can enter; we choose to
truncate at fourth order.  Symmetry considerations then require that it
take the form
\begin{equation}
E({\bf u}_i) = \kappa_2 u_i^2 + \alpha u_i^4
 + \gamma ( u_{ix}^2u_{iy}^2 + u_{iy}^2u_{iz}^2 + u_{iz}^2u_{ix}^2 )
\end{equation}
where $u_i = |{\bf u}_i|$, and $\kappa_2$, $\alpha$, and $\gamma$ are
expansion parameters to be determined from first-principles
calculations.

\subsection{Dipole-dipole interaction}

The second term in the effective Hamiltonian represents long-range
interactions between local modes.  Only dipole-dipole interactions
are considered, since higher-order terms tend to be of short range
and their effect will be included in the short-range contribution
$E^{\rm short} (\{ {\bf u}  \})$.  The dipole moment associated with
the local mode in cell $i$ is ${\bf d}_i= Z^* {\bf u}_i $.  Here, $Z^*$
is the Born effective charge for the soft mode, which can be
obtained as
\begin{equation}
Z^*=\xi_A Z^*_A+\xi_B Z^*_B
+\xi_{{\rm O}\parallel} Z^*_{{\rm O}\parallel}
+2\xi_{{\rm O}\perp} Z^*_{{\rm O}\perp}
\label{effcharge}
\end{equation}
from the eigenvector of the soft mode, once the Born effective
charges for the ions are known.\cite{zhong}
In atomic units (energy in Hartree), we have
\begin{equation}
E^{\rm dpl} (\{ {\bf u} \} ) = \frac{Z^{*2}}{\epsilon_\infty}
\sum_{i < j} \frac{{\bf u}_i
\cdot {\bf u}_j - 3 ( \hat{\bf R}_{ij} \cdot {\bf u}_i )
( \hat{\bf R}_{ij} \cdot {\bf u}_j ) } {R^3_{ij}} \; .
\label{dipolesum}
\end{equation}
Here, $\epsilon_{\infty}$ is the optical dielectric constant of the
material,
$R_{ij}=|{\bf R}_{ij}|$, ${\bf R}_{ij}= {\bf R}_i-{\bf R}_j$,
and $\hat{\bf R}_{ij} = {\bf R}_{ij}/ R_{ij} $.

In practice, Eq.~(\ref{dipolesum}) is not directly useful for
three-dimensional systems with periodic boundary conditions; instead,
we use an Ewald construction to evaluate $E^{\rm dpl}$.
We effectively terminate the sum in such a way that the $k=0$ modes
of the supercell will represent physical TO($\Gamma$) modes.  For
a TO mode, the induced depolarization electric field is zero;
from the point of view of the dipole sum, it is as though
the material were surrounded by a layer of metal.  In the
Ewald construction, this is equivalent to setting the surface terms
to zero.\cite{allen}  Under these conditions, and choosing the
decay $\lambda$ of the Gaussian charge packets to be small enough
so that the real-space summation can be entirely neglected, we have
\begin{equation}
E^{\rm dpl} = \frac{2 Z^{*2}}{\epsilon_{\infty}}
\left[ \frac{ \pi}{\Omega_c}
  \sum_{{\bf G}\ne 0} \frac{1}{|{\bf G}|^2} \exp
  \left( -\frac{|{\bf G}|^2}{4 \lambda^2} \right) \sum_{ij}
  ({\bf G}\cdot {\bf u}_i) ({\bf G}\cdot {\bf u}_j)
  \cos ( {\bf G}\cdot {\bf R}_{ij} )
  -\sum_i \frac{\lambda^3 u^2_i}{3 \sqrt{\pi}} \right] \; ,
\end{equation}
where $\Omega_c$ is the cell volume and {\bf G} is the reciprocal
lattice vector.

Because of its long-range nature, the calculation of $E^{\rm dpl}$ is
the most time-consuming part of our Monte Carlo simulations.  It
is thus worth some special treatment to reduce the computational
load.  In principle, the term $R_{ij}$ appearing in the denominator
of Eq.~(\ref{dipolesum}) should be strain-dependent. However, as we
have chosen to expand the intersite interactions between local
modes only up to harmonic order, it is consistent to ignore this
effect, since strain-induced changes of the dipole-dipole
interaction will enter only at higher order. This is equivalent to
fixing the reciprocal lattice vectors ${\bf G}$ and all the atomic
position vectors ${\bf R}_i$. The dipole energy can then be written
as
\begin{equation}
E^{\rm dpl} = \sum_{ij,\alpha\beta} Q_{ij,\alpha\beta} u_{i,\alpha}
u_{j,\beta},
\label{eqdpl}
\end{equation}
with
\begin{equation}
Q_{ij,\alpha\beta} = \frac{2 Z^{*2}}{\epsilon_{\infty}} \left[
\frac{\pi}{\Omega_c} \sum_{{\bf G}\ne 0}
\frac{1}{|{\bf G}|^2} \exp \left( -\frac{|{\bf G}|^2}{4 \lambda^2}
\right)
\cos ( {\bf G}\cdot {\bf R}_{ij}) G_{\alpha} G_{\beta}
  - \frac{\lambda^3 }{3 \sqrt{\pi}} \delta_{\alpha\beta} \delta_{ij}
\right] \; .
\end{equation}
Here, $\alpha$ and $\beta$ denote Cartesian components.  The matrix
$Q$ is thus treated as a constant; it is calculated once and for
all, and stored for later calculation of the dipole energy.  This
strategy increases the computational efficiency by at least one
order of magnitude.

\subsection{Short-range interaction}

$E^{\rm short} (\{ {\bf u}\})$ is the energy contribution due to
the short-range interactions between neighboring local modes, with
dipole-dipole interaction excluded. This contribution stems from
differences of short-range repulsion and electronic hybridization
between two adjacent local modes and two isolated local modes.
Together with the dipole-dipole interaction, this interaction
determines the soft-mode energy away from the zone center. Expanded
up to the second order, it can be written as,
\begin{equation}
E^{\rm short} (\{ {\bf u}\} ) = \frac{1}{2}
\sum_{i\ne j} \sum_{\alpha\beta}
J_{ij,\alpha\beta} u_{i\alpha} u_{j\beta} \; .
\label{eqshort}
\end{equation}
The coupling matrix $J_{ij,\alpha\beta}$ is a function of ${\bf
R}_{ij}$ and should decay very fast with increasing $|{\bf
R}_{ij}|$.  Here, we consider the short-range interaction up to
third nearest neighbor (nn), whose local mode shares atoms with the
local mode on the origin.  Local modes between further neighbors
involve displacements of atoms at least two hops away (in
tight-binding language) and their core-core repulsion or
hybridization should be much less important than the dipole-dipole
interaction which is taken care of in $E^{\rm dpl}$.

The interaction matrix $J_{ij,\alpha\beta}$ can be greatly simplified
by symmetry. For a cubic lattice, we have,
\begin{eqnarray}
{\rm 1st \; nn: \quad} & J_{ij,\alpha\beta} & =
         (j_1 +(j_2-j_1)|\hat{R}_{ij,\alpha}|)
   \delta_{\alpha\beta} \; , \nonumber\\
{\rm 2nd \; nn: \quad} & J_{ij,\alpha\beta}  & =
       (j_4 +\sqrt{2} (j_3-j_4) |\hat{R}_{ij,\alpha}|)
\delta_{\alpha\beta}
 \nonumber\\
   & & \qquad + 2 j_5 \hat{R}_{ij,\alpha} \hat{R}_{ij,\beta}
(1-\delta_{\alpha\beta})
   \; , \nonumber\\
{\rm 3rd \; nn: \quad} & J_{ij,\alpha\beta} & =
 j_6 \delta_{\alpha\beta} +
 3 j_7 \hat{R}_{ij,\alpha} \hat{R}_{ij,\beta}
(1-\delta_{\alpha\beta}) \; ,
\end{eqnarray}
where $\hat{R}_{ij,\alpha }$ is the $\alpha$ component of ${\bf
R}_{ij}/R_{ij}$.  So we have only seven interaction parameters for
a cubic lattice.  The coefficients $j_1$, $j_2$, ..., $j_7$ in the
above equations have physical meanings that are sketched
schematically in Fig.~{\ref{ji}}.  For example, $j_1$ represents
the interaction strength of ``$\pi$'' like interactions between
first neighbors.

\subsection{Elastic energy}

We will describe the state of elastic deformation of the BaTiO$_3$
crystal using local strain variables $\eta_l({\bf R}_i)$, where
the Voigt convention is used ($l=1,...,6$) and ${\bf R}_i$ labels
a cell center (Ti) site.  In fact, the six variables per unit cell
$\{\eta_l({\bf R}_i)\}$ are not independent, but are obtained from
three independent displacement variables per unit cell.  In our
analysis, these are taken as the dimensionless displacements
${\bf v}({\bf R}_i)$ (in units of the lattice constant $a$)
defined at the
unit cell corner (Ba) positions ${\bf R}_i+(a/2,a/2,a/2)$.
In terms of these, the {\it inhomogeneous} strain variables
$\eta_{{\rm I},l}({\bf R}_i)$ are defined in the next subsection.
Because of our use of a periodic supercell in the Monte Carlo
simulations, however, homogeneous strain deformations are not
included in the configuration space $\{ {\bf v}({\bf R}_i)\}$.
Therefore, we introduce six additional {\it homogeneous} strain
components $\eta_{{\rm H},l}$ to allow the simulation cell
to vary in shape.  The total elastic energy is expanded to
quadratic order as
\begin{equation}
E^{\rm elas}(\{ \eta_l \})=E_{\rm I}^{\rm elas}(\{ \eta_{{\rm I},l} \})
+E_{\rm H}^{\rm elas}(\{ \eta_{{\rm H},l} \}) \;,
\label{eqstrain}
\end{equation}
where the homogeneous strain energy is simply given by
\begin{eqnarray}
E^{\rm elas}_{\rm H}(\{\eta_{{\rm H},l}\}) & = &
\frac{N}{2} B_{11} (\eta_{{\rm H},1}^2 +
\eta_{{\rm H},2}^2 + \eta_{{\rm H},3}^2) \nonumber\\
 & + & N B_{12} (\eta_{{\rm H},1}\eta_{{\rm H},2} +
\eta_{{\rm H},2}\eta_{{\rm H},3} +
\eta_{{\rm H},3}\eta_{{\rm H},1}) \nonumber\\
 & + & \frac{N}{2} B_{44} (\eta_{{\rm H},4}^2 +
\eta_{{\rm H},5}^2 + \eta_{{\rm H},6}^2).
\label{eqhom}
\end{eqnarray}
Here $B_{11}$, $B_{12}$, and $B_{44}$ are the elastic constants
expressed in energy units ($B_{11}=a^3 C_{11}$, etc.),
and $N$ is the number of primitive cells in the supercell.

Rather than use an expression like (\ref{eqhom}) for the
inhomogeneous strain energy, we have found it preferable to express
this part directly in terms of the ${\bf v}({\bf R}_i)$.\cite{keating}
This approach keeps the acoustic phonon frequencies well behaved
throughout the Brillouin zone.
To satisfy requirements of invariance under translations and rotations
of the crystal as a whole, the energy is expanded in scalar products of
differences between the ${\bf v}({\bf R}_i)$.  The cubic crystal
symmetry leads to a great reduction of the independent parameters
in the expansion.  The energies of the long-wavelength strain
deformations can be reproduced by an expansion of the form
\begin{eqnarray}
E_{\rm I}^{\rm elas} &=& \sum_i \Big\{ \gamma_{11}
[v_x({\bf R}_i)-v_x({\bf R}_i \pm {\bf x})]^2 \nonumber\\
\phantom{\Big|}
  &+& \gamma_{12}
[v_x({\bf R}_i) - v_x({\bf R}_i \pm {\bf x} )]
[v_y({\bf R}_i) - v_y({\bf R}_i \pm {\bf y} )] \nonumber\\
\phantom{\Big|}
  &+& \gamma_{44}
[v_x({\bf R}_i) - v_x({\bf R}_i  \pm {\bf y} ) +
v_y({\bf R}_i) - v_y({\bf R}_i  \pm {\bf x} )]^2 \nonumber\\
  &+& {\rm \;cyclic\;permutations\;} \Big\}
\label{eqinhom}
\end{eqnarray}
corresponding to bond stretching, bond correlation, and bond bending,
respectively.  Here, $ {\bf x} \equiv a\hat{\bf x}$,
${\bf y} \equiv a\hat{\bf y}$, $ {\bf z} \equiv a\hat{\bf z}$,
and $\pm$ indicates multiple terms to be summed.
The $\gamma$ coefficients are related to the elastic constants
by $\gamma_{11} = B_{11}/4$, $\gamma_{12}= B_{12}/8$, and
$\gamma_{44}=B_{44}/8 $.

\subsection{Elastic-mode interaction}

To describe the coupling between the elastic deformations and the
local modes, we use the on-site interaction
\begin{eqnarray}
E^{\rm int} (\{ {\bf u}\}, \{\eta_l \}) =
{1\over 2}\sum_i\sum_{l\alpha\beta}B_{l\alpha\beta}\eta_l ({\bf
R}_i)u_\alpha({\bf R_i})u_\beta({\bf R_i}) \;.
\label{eqelint}
\end{eqnarray}
As a result of cubic symmetry, there are only three independent
coupling constants $B_{l\alpha\beta}$:
\begin{eqnarray}
&&B_{1xx} = B_{2yy} = B_{3zz} \;,\nonumber\\
B_{1yy} = &&B_{1zz} = B_{2xx}= B_{2zz} = B_{3xx} = B_{3yy}\;,\nonumber\\
B_{4yz} = &&B_{4zy} = B_{5xz}= B_{5zx} = B_{6xy} = B_{6yx}\;.\nonumber
\end{eqnarray}
The strain contains both homogeneous and inhomogeneous parts.
$\eta_l({\bf R}_i)=\eta_{{\rm H},l}({\bf R}_i)+
\eta_{{\rm I},l}({\bf R}_i)$.
The latter are expressed in terms of
the local displacement vectors ${\bf v}$ as follows.
We first define the six average differential
displacements associated with site ${\bf R}_i$ as
\begin{eqnarray}
 \Delta v_{xx} & = &
\sum_{{\bf d} = 0,{\bf y},{\bf z},{\bf y}+{\bf z}}
\left[ v_x ({\bf R}_i - {\bf d} - {\bf x}) - v_x ({\bf
 R}_i - {\bf d} ) \right] \;,\nonumber\\
 \Delta v_{xy} & = &
\sum_{{\bf d} = 0,{\bf y},{\bf z},{\bf
y}+{\bf z}} \left[ v_y ({\bf R}_i - {\bf d} - {\bf x}) - v_y ({\bf
R}_i - {\bf d} ) \right] \;,\nonumber
\end{eqnarray}
and their cyclic permutations, recalling
that ${\bf v}({\bf R}_i)$ is associated with position
${\bf R}_i+(a/2,a/2,a/2)$.  Then
$\eta_{{\rm I},1} ({\bf R}_i)= \Delta v_{xx} / 4$,
$\eta_{{\rm I},4} ({\bf R}_i)= (\Delta v_{yz} +
\Delta v_{zy})/ 4$, etc.

\section{First-principles calculations}

We have shown that, with the two approximations we made, the total
energy functional of the perovskite system is fully specified by a set
of parameters.  These parameters can be obtained from first-principles
calculations. We use density-functional theory within the local density
approximation (LDA). The technical details and convergence tests of the
calculations can be found in Refs. \onlinecite{king1,king11}.  The most
important feature of the calculations is the Vanderbilt ultra-soft
pseudopotential,\cite{vand1} which allows a low energy cutoff to be
used for first-row elements.  This makes high-accuracy large-scale
calculations of materials involving oxygen and $3d$ transition metal
elements affordable.  The ultra-soft pseudopotential also allows for
exceptionally transferable pseudopotentials. It ensures that
all-electron and pseudo-atom scattering properties agree over a very
large energy range and preserves the chemical hardness of the atom.  A
generalized Kohn-Sham functional is directly minimized using a
preconditioned conjugate gradient method.\cite{king11,payne,aria}  We
use a (6,6,6) Monkhorst-Pack k-point mesh\cite{monk} for single-cell
calculations, i.e., 216 k-points in the full Brillouin zone (FBZ).  For
supercell calculations, the k mesh is kept the same to minimize errors
due to the k-point sampling.  Therefore, a smaller number of k-points
is used because of the smaller FBZ.

We start with the construction of the local mode vectors.  All the
eigenvalues and eigenvectors of the force-constant matrix at $k=0$ for
the cubic BaTiO$_3$ structure are calculated from frozen-phonon
calculations, as in Ref. \onlinecite{king11}. The mode with imaginary
frequency is identified as the soft mode.  The soft-mode eigenvector
has been reported previously\cite{king11} as $\xi_{\rm Ba} = 0.20$,
$\xi_{\rm Ti} =0.76$, $\xi_{{\rm O}\parallel} = -0.53$, and $\xi_{{\rm
O}\perp} = -0.21 $.  The local mode
is then constructed from it using the scheme proposed in II.1
(Fig.\ref{stru}).

Determination of many of the parameters in the
effective Hamiltonian involves
only calculations for zone-center distortions.  These parameters have
been reported previously.\cite{king11,zhong} They include the
fourth-order terms of on-site energy $\alpha$ and $\gamma$; the elastic
constants $B_{11}, B_{12}, B_{44}$; and the on-site elastic-mode
interaction parameters $B_{1xx}, B_{1yy}, B_{4yz}$.  The mode
effective charge $Z^*$ of Eq.\ (\ref{effcharge}) is calculated from
the values
$Z^*_A$=2.75, $Z^*_B$=7.16
$Z^*_{{\rm O}\parallel}$=$-$5.69, and
$Z^*_{{\rm O}\perp}$=$-$2.11 published in Ref.\ \onlinecite{zhong}.
(The resulting value $Z^*$=9.96 is slightly different from the one
given in Ref.\ \onlinecite{zhong}; there, the eigenvector of the
{\it dynamical} matrix, not the {\it force-constant} matrix,
was used.) We use the experimental value $\epsilon_\infty=5.24$
of the optical dielectric constant, since for this quantity, the LDA
seems not to be a well-justified approximation, while exact
density-functional theory results are not
accessible.  We find, however, that the effect of a small inaccuracy in the
dielectric constant affects thermodynamic properties such as transition
temperatures only slightly.

The second-order energy parameter $\kappa$ for zone-center distortions
is a linear combination of the local mode self energy parameter
$\kappa_2$, the intersite interactions $j_i$, and the dipole-dipole
interaction.  The calculations of intersite interaction parameters
involve determination of the energy for distortions at the
zone-boundary k-points X=($\pi/a$,0,0), M=($\pi/a$,$\pi/a$,0), and
R=($\pi/a$,$\pi/a$,$\pi/a$), where $a$ is the lattice constant.  Five
frozen-phonon calculations on doubled unit cells are sufficient to
extract all the information available from these k-points.  The
arrangement of the local mode vectors for each case, as well as for the
zone-center distortion at $\Gamma$=(0,0,0), are shown in
Fig.~\ref{cell}.  The actual ionic configurations are constructed by
superpositions of displacements associated with adjacent local modes.
For example, letting $u$ be the amplitude of the Ti-centered local mode
defined in Sec.\ \ref{sec:approx}, the displacement of the Ti atoms is
just $u\xi_{\rm Ti}$ in Fig.\ \ref{cell}(a) and $\pm u\xi_{\rm Ti}$ in
Fig.\ \ref{cell}(b), while Ba is affected by eight neighboring local
modes so that its displacement is $8\times u\xi_{\rm Ba}/8=u\xi_{\rm
Ba}$ in Fig.\ \ref{cell}(a) and $0$ in Fig.\ \ref{cell}(b).

The above five doubled-cell calculations can be used to determine the
parameters $j_1$, $j_2$, $j_3$, $j_4$, and $j_6$.  The determination of
$j_5 + 2 j_7 $ requires a four-cell calculation involving 20
atoms with low symmetry [Fig.~\ref{cell}(g)].  Table~\ref{table1}
lists the energy expressions for all the configurations calculated in
terms of the quadratic expansion parameters.

A further decomposition of $j_5$ and $j_7$ would require an expensive
eight-cell calculation. Furthermore, the interaction parameter $j_7$ is
the third nearest-neighbor interaction and is thus presumably not very
important.  This argument is justified by our Monte Carlo simulations
which show that the calculated transition temperature is insensitive to
different decompositions of $j_5$ and $j_7$. This prompts us to make an
approximate decomposition based on a simple physical argument:  we
expect the interaction to be smallest for two interacting local modes
oriented such that reversing the relative sign of the vectors produces
the least change of bond lengths.  Applied to third nearest neighbors,
this argument implies $j_6 - 2j_7 =0$, thus fixing the value of $j_7$.

The resulting interaction parameters are shown in Table~\ref{table2},
together with other parameters published previously.  It may be
surprising to see that the on-site $\kappa_2$ is positive, while the
cubic structure is known to be unstable against $k=0$ distortion.
The cubic structure is thus stable against forming an isolated local
mode; instability actually comes from the intersite interactions
between local modes.  To be more precise, we find that it is
the long-range Coulomb (dipole-dipole) interaction which
makes the ferroelectric state favorable.  If we turn off the
dipole-dipole interaction by setting the effective charge $Z^* = 0$,
we find that the ferroelectric instability disappears.  This is
consistent with the previous point of view that long-range Coulomb
forces favor the ferroelectric state, while short-range repulsions
favor the nonpolar cubic state.

{}From Table~\ref{table2}, we see that the intersite interaction
parameters decay very fast with increasing distance, indicating
the short-range nature of the intersite interactions after the
long-range Coulomb interactions have been separated out.
The ratio of the magnitudes of the strongest first-, second-, and
third-neighbor interactions turns out to be approximately
$ 1 : 0.23 : 0.09 $.  This decays even faster than the
dipole-dipole interactions, for which the corresponding ratio
($\propto 1/R^3$) is $ 1 : 0.35 : 0.19 $.
These results help justify our approximation of including only
up to third nearest neighbors for the short-range interactions.

\section{Monte Carlo simulations}

For the quantitative study of non-universal finite-temperature behavior
of a given model, Monte Carlo simulation\cite{allen,binder1} has
emerged as the most reliable and powerful technique.  It is especially
appropriate for a model such as ours, with two continuous vector
degrees of freedom per unit cell and both short and long range
interactions, for which analytical approaches such as renormalization
group or high-temperature expansions would be cumbersome and involve
additional approximations.  In comparison, the Monte Carlo approach
requires only the ability to compute changes in total energy as the
configuration is changed.  Furthermore, with suitable analysis of
statistical errors and finite-size effects, the results of Monte Carlo
simulation can be made arbitrarily accurate. Finally, with little
additional effort, a number of physical quantities can be computed to
aid in characterization of the transition.

We solve the effective Hamiltonian [Eqs.\ (\ref{eqself}),
(\ref{eqdpl}), (\ref{eqshort}), (\ref{eqstrain}), and (\ref{eqelint})]
using Monte Carlo simulations with the Metropolis algorithm\cite{metro}
on an $L\times L \times L$ cubic lattice with periodic boundary
conditions.  Since most energy contributions (except $E^{\rm dpl}$) are
local, we choose the single-flip algorithm.  That is, a trial move
consists of an attempted update of a single variable, after which the
total energy change is calculated to determine whether to accept the
move.  The step sizes are adjusted to ensure an acceptance ratio of
approximately 0.2.  In one Monte Carlo sweep (MCS), we first make a
trial move on each ${\bf u}_i$ in sequence, then each ${\bf v}_i$ in
sequence, then iterate several times (typically $2L$ times) on the
homogeneous strain variables.  For $L=12$, each MCS takes about one
second on an HP 735 workstation.  The typical correlation time for the
total energy is found to be several hundred MCS close to the phase
transition; this long correlation time makes certain new MC techniques
using energy distribution functions\cite{ferr} unfavorable.  The
correlation times for the local mode amplitudes are one order of
magnitude shorter, and thus 10,000 MCS are usually enough to
equilibrate and to obtain averages of local mode variables with
statistical error $< 10$\%.

In our simulation, we concentrate on identifying the succession of
low-temperature phases, determining the phase transition temperatures,
and extracting qualitative features of the transitions. This analysis
will allow us to identify the features of the effective Hamiltonian which
most strongly affect the transition properties.  For these purposes, it
is most convenient to monitor directly the behavior of the homogeneous
strain and the vector order parameter.  In the case of the
ferroelectric phase transition, the latter is just the average
local-mode vector ${\bf u} = \sum_i {\bf u}_i/ N$, which is
proportional to the polarization.  Here, ${\bf u}_i$ is the local mode
vector at site $i$ and $N$ is the total number of sites. As a
reference, the average local mode amplitude $u = \sum_i |{\bf u}_i| /
N$ is also monitored. To avoid effects of symmetry-equivalent rotations
of the order parameter and to identify the different phases clearly, we
accumulate the absolute values of the largest, middle, and smallest
components of the averaged local-mode vector for each step, denoted by
$u_1$, $u_2$, and $u_3$, respectively ($u_1 > u_2 > u_3$).  The cubic
(C), tetragonal (T), orthorhombic (O), and rhombohedral (R) phases are
then characterized by zero, one, two, and three non-zero
order-parameter components, respectively.  The effect of
symmetry-equivalent rotations on the homogeneous strain is handled
analogously, with the largest, the medium and the smallest linear
strain components denoted by $\eta_1$, $\eta_2$ and $\eta_3$,
respectively, and shear strain components by $\eta_4$, $\eta_5$ and
$\eta_6$.

The ferroelectric phase transition is very sensitive to hydrostatic
pressure, or equivalently, to lattice constant.  The LDA-calculated
lattice constants are typically 1\% too small, and even this small
error can lead to large errors in the zero-pressure transition
temperatures.  One approach which largely compensates for the effect of
this systematic error is to exert a negative pressure that expands the
lattice constant to the experimental value.  We determine the value of
the pressure by calculating volumes for four different phases and
comparing with experimental measurements.\cite{lb}  We find $P=-4.8$
GPa gives the best overall agreement (although the application of
pressure does lead to a slight change in the low temperature
structure). Except for the simulations for the construction of the
temperature-pressure phase diagram, the following simulations and
analysis are for this pressure.

\section{Results and Discussion}

In this section, the finite-temperature behavior of the model is
presented and analyzed. First, we examine the order parameters as a
function of temperature in a typical simulation to obtain a measure of
the transition temperatures.  {}From the results of simulations for a
range of pressures, we construct the temperature-pressure phase
diagram.  For the system at ambient pressure, more detailed simulations
are performed.  The order of the transitions, the nature of the
paraelectric phase and the properties of the low-temperature phases are
investigated and compared with experimental observations.  Finally, the
roles played by different terms in the effective Hamiltonian and the
sensitivity of the results to various approximations are examined.

\subsection{Order parameters and phase diagram}

We start the simulations at a high temperature ($T>400$K)
and equilibrate for 10,000 MCS.  The temperature is then reduced in small
steps, typically 10K.  After each step, the system is allowed to
equilibrate for 10,000 MCS.  The order parameter averages are then
accumulated over a period of 7,000 MCS, yielding a typical standard
deviation of less than 10\%.
The temperature step size is reduced and the number of MCS
is doubled for temperatures close to the phase transition.

Fig.~\ref{u-T} shows the averaged local-mode vectors $u_1$, $u_2$,
$u_3$ and averaged local mode amplitude $u$ as functions of temperature
in a typical simulation for an $L=12$ lattice at $P=-4.8$ GPa.
At high
temperatures, $u_1$, $u_2$, and $u_3$ are all very close to zero.  As
the system is cooled down below 295K, $u_1$ increases and becomes
significantly larger than $u_2$ or $u_3$.  This indicates the
transition to the tetragonal phase.  Two additional phase transitions
occur as the temperature is reduced further.  The T--O transition
(sudden increase of $u_2$) occurs at 230K  and the O--R transition
(sudden increase of $u_3$) occurs at 190K.  The sequence of transitions
exhibited by the simulation is the same as that observed
experimentally.

The averaged homogeneous strain variables obtained from the above
simulation are shown in Fig.~\ref{e-T}.  These strains are measured
relative to the LDA calculated equilibrium cubic structure, so the
linear strains are significantly non-zero at higher temperatures due to
the negative pressure applied.  As expected, the simulation cell
changes shape at the same temperatures at which the jumps of order
parameter components are observed.  At high temperatures,
we have approximately
$\eta_1=\eta_2=\eta_3$ and $\eta_4=\eta_5=\eta_6=0$, corresponding to
the cubic structure.  As the system is cooled down, the shape of the
simulation cell changes to T, O, and R phases.  The orthorhombic (O)
structure has a non-zero shear strain, in agreement with the centered
orthorhombic structure observed experimentally.  Quantitatively, our
calculated distortions are also in good agreement with the experiment,
with the calculated distortions tending to be slightly smaller.  For
example, $\eta_1-\eta_3$ for the tetragonal phase is 1.1\% as measured
from experiment\cite{lb} and 0.9\% from our calculation.

The simulations are repeated for a range of applied pressures to obtain
the temperatures at which the order parameter components and
homogeneous strain jump on cooling down.  The resulting
temperature-pressure phase diagram is shown in Fig.\ \ref{p-T}.  (This
measure of the transition temperature is actually a lower bound, due to
hysteresis around $5\%$ for T--O and O--R transitions and negligible
for the C--T transition, to be discussed further below.) All three
transition temperatures decrease almost linearly with increasing
pressure.  At the experimental lattice constant, the values for
$dT_c/dP$ are found to be $-$28, $-$22, and $-$15 K/GPa for the C--T, T--O,
and O--R transitions respectively. The experimental values for the C--T
transition range from $-$40 K/GPa\cite{clarke} to $-$66 K/GPa.\cite{samara}
For the T--O transition the measured value is $-$28 K/GPa,\cite{samara}
and for the O--R transition it is $-$10 to $-$15 K/GPa.\cite{samara}
At pressures as high as $P=5$GPa, the sequence of phases
C--T--O--R is still observed in the simulation.
When the pressure is increased
further, the phase boundary in the simulation becomes unclear due to
fluctuations.  Our calculated critical pressure (beyond which the cubic
structure is stable at T=0 K) is $P_c=8.4$GPa.  Taking into account the
pressure correction for the LDA volume underestimate, this corresponds
to a predicted physical $P_c=13.2$ GPa.
We are not aware of any experimental value for $P_c$ with which
to compare this value.  However, we find that the magnitude of our
$dT_c/dP$ is significantly smaller than experimental value, at least
for the C--T and T--O transitions.
This may partly be due to the neglect of higher-order strain
coupling terms in the effective Hamiltonian.
We have tested the effect of including a volume dependence for the
short-range interaction parameters $j_i$.  This correction does
not change the sequence of phases, and it only increases the
magnitudes of
$dT_c/dP$ slightly.  Therefore, our results are reported
without this correction.  The accuracy of the phase
diagram may be further improved by including higher-order terms in
the elastic energy, or the coupling of $j_i$ to anisotropic strain.

\subsection{Hysteresis and latent heat}

For the investigation of the order of the transitions, the nature of
the paraelectric phase and the properties of the low-temperature
phases, we performed more detailed simulations at  $P=-4.8$GPa for
the three system sizes $L=10$, $12$ and $14$.  In the cooling-down
simulations, the length of each simulation was increased from 10,000 to
up to 35,000 MCS at temperatures close ($\pm$10K) to the phase
transition to include a longer equilibration.
The size of the temperature step was decreased to 5K or less in the
vicinity of the transition.  In addition, a heating simulation was
performed, starting from the lower-temperature phase, to detect
any possible hysteresis.
The calculated transition temperatures, obtained as the
average of the cooling and heating transition temperatures, are given
in Table \ref{table3}.  The error estimates in the table are determined
by the width of the hysteresis, which persists even for the longest
simulation lengths considered.  (The C--T transition temperature for
$L=10$ is difficult to identify because of large fluctuations between
phases.) The calculated transition temperatures are well converged with
respect to system size, and are in reasonable agreement with
experiment.  Table \ref{table3} also gives the saturated spontaneous
polarization p$_{\rm s}$ at $T=0$ in the R phase, and just above the
O--R and T--O transitions in the O and T phases respectively.
These are calculated from the average local mode vector ${\bf u}$ and the
local mode $Z^*$.  We find almost no finite-size effect for this
quantity, and the agreement with experiment is very good for the O
and T phases.  The disagreement for the R phase may result in part
from twinning effects in the experimental sample.\cite{wied}

{}From the jumps in structural parameters and the observed hysteresis
in heating and cooling, we conclude that the phase transitions are
first order.  An accurate
determination of the latent heats would require considerable
effort;\cite{janke}  here, we only try to provide estimates
sufficiently accurate for meaningful comparison with experiment.  We
approach each transition from both high-temperature and low-temperature
sides until the point is reached where both phases appear equally
stable.
(That is, the typical time for the system to fluctuate into
the opposite phase is roughly independent of which phase the
simulation is started in.)
The difference of the average total energy
is then the latent heat.\cite{latent} This approach is practical as
long as some hysteresis is present.  The calculated latent heats (Table
\ref{table3}) show non-negligible finite-size dependence. Taking this
into account, we find that the latent heats for all three transitions
are significantly non-zero and in rough agreement with the rather
scattered experimental data.  For the T--O and O--R transitions, the
first-order character of the transition is predicted by Landau theory,
since in these two cases the symmetry group of the low temperature
structure is not a subgroup of that of the high temperature structure.
For the C--T transition, the first-order character is not a consequence
of symmetry, but rather of the values of the effective Hamiltonian
parameters.  Although it has the largest latent heat of the three
transitions, the C--T transition also exhibits large finite-size
effects in the latent heat and in the smearing of order parameter
components and strain discontinuities in the simulation
(Figs.~\ref{u-T},\ref{e-T}). This suggests the presence of long-wavelength
fluctuations characteristic of second-order phase transitions and
critical phenomena, and the classification of the C--T transition as a
weak first-order transition.

\subsection{Displacive vs.\ order-disorder}

Using the microscopic information available in the simulations, we are
able to investigate the extent to which the cubic-tetragonal transition
can be characterized as order-disorder or displacive.  These possibilities
can be distinguished by inspecting the distribution of the real-space
local-mode vector ${\bf u}_i$ in the cubic phase just above the
transition.  A displacive (microscopically nonpolar) or order-disorder
(microscopically polar) transition should be characterized by a
single-peaked or double-peaked structure, respectively.  The
distribution of $u_x$ at $T=400$K is shown in Fig.~\ref{udos}.  It
exhibits a rather weak tendency to a double-peaked structure,
indicating a transition which has some degree of order-disorder
character.  We also see indications of this in the $u$--$T$ relation in
Fig.~\ref{u-T}. Even in the cubic phase, the average of the local-mode
magnitude $u$ is significantly non-zero and close to that of the
rhombohedral phase, while the magnitudes of the average local mode
components change dramatically during the phase transitions.

In reciprocal space, a system close to a displacive transition should
show large and strongly temperature-dependent fluctuations of certain
modes associated with a small portion of the Brillouin zone (BZ) (for a
ferroelectric transition, near $\Gamma$).  For an extreme
order-disorder transition, on the other hand, the fluctuations are
expected to be distributed over the whole BZ.  For BaTiO$_3$, we
calculated the average Fourier modulus $F({\bf k},T) = \langle|u({\bf
k})|^2\rangle$ for eigenmodes at several high-symmetry k-points (along
$\Gamma$--X, $\Gamma$--M and $\Gamma$--R) for a range of temperatures
above the C--T transition.  These eigenmodes are identified by their
symmetry properties as one longitudinal optical (LO) branch and two
transverse optical (TO) branches at each point.  For a purely harmonic
system, $ T / F({\bf k},T) $ can be shown to be a
temperature-independent constant proportional to the square of the
eigenfrequency $\omega^2({\bf k})$ of the corresponding eigenmode.  A
strong decrease of $T / F({\bf k},T)$ as $T\rightarrow T_c$ from high
temperature can be interpreted as mode softening due to anharmonicity.

The results at the k-point ($\pi/4a$, $\pi/4a$, $0$) illustrate the
main features of this analysis. From symmetry, three nondegenerate
eigenmodes LO, TO1, TO2 are identified. The polarization of LO, TO1,
and TO2 are in the direction of $\hat{\bf x} + \hat{\bf y}$, $\hat{\bf
x}-\hat{\bf y}$, and $\hat{\bf z}$, respectively.  For each eigenmode,
the temperature dependence of the calculated $\omega^2({\bf k},T)$ is
shown in Fig.~\ref{u2g}.  The almost linear behavior of $\omega^2({\bf
k},T)$ vs.\ $T$ (the Curie-Weiss form) is observed for the other
k-points as well.  Both the LO and TO1 branches are almost temperature
independent. The TO2 branch is strongly temperature dependent and is
thus a ``soft'' mode.  According to the soft-mode theory of structural
phase transitions, $T_c$ is the lowest temperature at which all
$\omega^2({\bf k},T)\ge 0$.  Linear extrapolation indicates that the
TO2 mode frequency goes to zero at $T\approx 200$K, which is a lower
bound for $T_c$, consistent with the value obtained in Monte Carlo
simulations.  A similar calculation of $\omega^2({\bf k},T)$ for the TO
modes at $\Gamma$=(0,0,0) extrapolates to zero at the higher
temperature $T\approx 300$K, in excellent agreement with the Monte
Carlo value of $295$K.

Within this formalism, the microscopic character of the paraelectric
phase is determined by the extent of the soft mode in the BZ.  We
define a quantity
\begin{equation}
\rho ({\bf k}) = \frac{2 F({\bf k},350{\rm K})}{F({\bf k},700{\rm K})}
\label{rhodef}
\end{equation}
to indicate the hardness of the modes. In Fig.~\ref{u2g2}, $\rho ({\bf
k})$ is shown for the various k-points along some special directions in
the BZ.  If $\rho ({\bf k}) < 1$, the corresponding eigenfrequency
extrapolates to zero at some positive temperature, and the mode is
regarded as soft.  If $\omega^2({\bf k})$ is independent of
temperature, $\rho ({\bf k}) =2$, corresponding to the hardest mode.

For all the k-points considered, all the LO modes are almost
temperature independent [$\rho ({\bf k}) =2$] and are not included in the
figure.  Along the $\Gamma$--X direction, the doubly-degenerate TO
modes are soft all the way to the zone boundary. In contrast, along the
$\Gamma$--R direction, both TO modes become hard immediately after
leaving the $\Gamma$ point.  Along the $\Gamma$--M direction, the TO1 mode
becomes hard quickly, while the TO2 branch remains soft at least
halfway to the zone boundary.  This behavior, especially along
$\Gamma$--X, does not conform completely to the displacive limit. This
supports the interpretation of the C--T transition as intermediate
between displacive and order-disorder, with a slightly stronger
order-disorder character.  Thus, from the example of BaTiO$_3$, it
seems that a positive on-site quadratic coefficient does not
automatically imply a displacive character for the transition.  Rather,
the relevant criterion is the extent to which the unstable modes
extend throughout the BZ.

\subsection{Roles of different interactions}

Our theoretical approach allows us to investigate the roles played by
different types of interaction in the phase transition.  First, we
study the effect of strain.  Recall that the strain degrees of freedom
were separated into local and homogeneous parts, representing finite-
and infinite-wavelength acoustic modes, respectively.  Both parts were
included in the simulations.  If we eliminate the local strain (while
still allowing homogeneous strain), we find almost no change in the
transition temperatures.  This indicates that the effect of the
short-wavelength acoustic modes may not be important for the
ferroelectric phase transition.  If the homogeneous strain is frozen at
zero, however, we find a direct
cubic--rhombohedral phase transition, instead of the correct series of
three transitions.  This demonstrates the important role of homogeneous
strain.

Second, we studied the significance of the long-range Coulomb
interaction in the simulation. To do this, we changed the effective
charge of the local mode (and thus the dipole-dipole interaction
strength), while modifying the second-order self-energy parameter
$\kappa_2$, so that the frequencies of the zone-center and
zone-boundary modes remain in agreement with the LDA values.  We
found only a slight change (10\%) of the transition temperatures when
the dipole-dipole interaction strength was doubled.  However, elimination
of the dipole-dipole interaction altogether changed the results
dramatically; the ground state becomes a complex antiferroelectric
structure similar to the room-temperature structure of PbZrO$_3$. This
result shows that it is essential to include the long-range
interaction, although small inaccuracies in the calculated values of
the effective charges or dielectric constants may not be very
critical.

Third, we investigated the sensitivity of our results to variations
of the short-range interaction parameters.  We find the accuracy of
the first-neighbor
interaction parameters ($j_1$,$j_2$) is very important, and a
mere 10\% deviation can change the calculated transition temperatures
dramatically, and can sometimes even change the ground state structure.
Second nearest-neighbor interactions are less important, and for the
third-neighbor interactions, even a 100\% change does not
seem to have a strong effect on the values of T$_c$.  This result
is to be expected, and partly justifies our choice of including
only up to third neighbors for the short range interactions.
We have also tested the effect of
our assumption $j_6-2j_7=0$. We find that any reasonable
choice leads to a barely noticeable change in $T_c$.

In short, highly accurate LDA calculations do appear to be a
prerequisite for an accurate determination of the transition
temperatures, but as long as certain features of the energy surface are
correctly described, other approximations can be made without
significantly affecting the results.

\section{Conclusions}

We have developed a first-principles approach to the study of structural
phase transitions and finite-temperature properties in perovskite
compounds.  We construct an effective Hamiltonian based on Taylor
expansion of the energy surface around the cubic structure,
including soft optical modes and strain components as the
possible distortions.  The expansion parameters are determined by
first-principles density-functional calculations using Vanderbilt's
ultra-soft pseudopotential.

We have applied this scheme to BaTiO$_3$ and calculated the
pressure-temperature phase diagram. We have obtained the sequence of
low-temperature phases, the transition temperatures, and the spontaneous
polarizations, and found them to be in good agreement with experiment.
We find that long-wavelength acoustic modes and long-range dipolar
interactions both play an important role in the phase transition, while
short-wavelength acoustic modes are not as significant. Accurate LDA
calculations are required for accurate determination of the transition
temperatures.  The C--T phase transition is not found to be well
described as a simple displacive transition; on the contrary, if
anything it has more order-disorder character.

With slight modifications, our approach should be applicable to
other perovskite compounds, as long as their structure is close to
cubic and there are some low energy distortions responsible for
the phase transitions.  It can be easily applied to ferroelectric
materials like PbTiO$_3$ (under study by another
group\cite{rabeuw}) and KNbO$_3$. It can also be applied to
antiferroelectric materials like PbZrO$_3$.  The application to
antiferrodistortive materials like SrTiO$_3$ is slightly more
involved, though also successful.\cite{zhong2}

\acknowledgments

We would like to thank R.D. King-Smith, U. V. Waghmare, R. Resta,
Z.~Cai, and A.M.~Ferrenberg for useful discussions.
This work was supported by the Office of Naval Research under contract
number N00014-91-J-1184 and N00014-91-J-1247.

\newpage

\begin{table}
\caption{The energy per 5-atom unit cell (excluding the dipole energy)
in terms of intersite interaction parameters of Fig.~2,
for configurations shown in Fig.~3.\label{table1}}
\begin{tabular}{clc}
Configuration & expression \\ \tableline
(a) & $ \kappa_2+2j_1 +  j_2 + 4 j_3 + 2j_4 + 4j_6 $ \\
(b) & $ \kappa_2+2j_1 -  j_2 - 4 j_3 + 2j_4 - 4j_6 $ \\
(c) & $ \kappa_2+  j_2 - 2j_4 - 4j_6 $ \\
(d) & $ \kappa_2-2j_1 +  j_2 - 4 j_3 + 2j_4 + 4j_6 $ \\
(e) & $ \kappa_2      +  j_2         - 2j_4 + 4j_6 $ \\
(f) & $ \kappa_2-2j_1 -  j_2 + 4 j_3 + 2j_4 - 4j_6 $ \\
(g) & $ \kappa_2/2+j_1 -2 j_5 - 4j_7 $
\end{tabular}
\end{table}

\begin{table}
\caption{Expansion parameters of the Hamiltonian for BaTiO$_3$.
Energies are in Hartrees.
\label{table2}}
\begin{tabular}{l|cr|cr|cr}
On-site & $\kappa_2$ & 0.0568 & $\alpha$ & 0.320 & $\gamma$ & $-$0.473
        \\\hline
        &  $j_1$ & $-$0.02734 & $j_2$ & 0.04020  \\
Intersite & $j_3$ & 0.00927 & $j_4$ & $-$0.00815 & $j_5$ & 0.00580 \\
           & $j_6$ & 0.00370 & $j_7$ & 0.00185 \\\hline
Elastic & $B_{11}$ & 4.64 & $B_{12}$ & 1.65 & $B_{44} $ & 1.85 \\\hline
Coupling & $B_{1xx}$ & $-$2.18 & $B_{1yy}$ & $-$0.20 & $B_{4yz} $ & $-$0.08
  \\\hline
Dipole & $Z^*$ & 9.956 & $\epsilon_{\infty}$ &5.24
\end{tabular}
\end{table}

\begin{table}
\caption{Calculated transition temperatures $T_{\rm c}$,
saturated spontaneous polarizations p$_{\rm s}$, and estimated
latent heats $l$, as a function of simulation cell size.
\label{table3}}
\begin{tabular}{l|c|cccc}
         &phase & $L=10$   & $L=12$ & $L=14$ & expt\tablenotemark[1]\\
\hline
                & O--R & 197$\pm$3 & 200$\pm$10 & 200$\pm$5 & 183 \\
$T_{\rm c}$ (K) & T--O & 230$\pm$10  & 232$\pm$2  & 230$\pm$10 & 278 \\
                & C--T & $\sim$290     & 296$\pm$1  & 297$\pm$1 & 403 \\
\hline
                & R    & 0.43       & 0.43       &  0.43     & 0.33 \\
p$_{\rm s}$ (C/m$^2$) & O & 0.35    & 0.35       &  0.35     & 0.36 \\
                & T   &     0.28    & 0.28       &  0.28     & 0.27 \\
\hline
%
                & O--R &  50        &    60      &  60     & 33--60 \\
$l$ (J/mol)     & T--O &  90    &    90      & 100       & 65--92 \\
                & C--T & --         &  --   & 150       & 196--209 \\
\end{tabular}
\tablenotetext[1]{Ref.~\onlinecite{lb}}

\end{table}
\newpage

\begin{figure}
\caption{The structure of the cubic perovskite compound BaTiO$_3$.
Ba, Ti and O atoms are represented by shaded, solid, and empty circles
respectively. Lengths of the vectors indicate the relative magnitudes of
the displacements for a local mode polarized along $\hat{\bf x}$.
 \label{stru} }
\end{figure}

\begin{figure}
\caption{The independent intersite interactions
corresponding to the parameters $j_1$, $j_2$ (first neighbor),
$j_3$, $j_4$, $j_5$ (second neighbor), and $j_6$ and $j_7$ (third neighbor).
\label{ji} }
\end{figure}

\begin{figure}
\caption{The local mode arrangements for which first-principles total-energy
calculations were performed to determine
the intersite interaction parameters. The arrangements can be labeled
by the wavevector {\bf k} and a polarization vector
($\hat{\bf p}$).
The arrows represent local mode vectors. The dotted lines
indicate the unit cells of the simple cubic lattice.
The solid lines show the supercells used in the calculations.
(a) $\Gamma$, $\hat{\bf p} = \hat{\bf z}$ ;
(b) $X$, $\hat{\bf p} = \hat{\bf z}$ ;
(c) $X$, $\hat{\bf p} = \hat{\bf x}$ ;
(d) $M$, $\hat{\bf p} = \hat{\bf z}$ ;
(e) $M$, $\hat{\bf p} = \hat{\bf x}$ ;
(f) $R$, $\hat{\bf p} = \hat{\bf z}$ ;
(g) four-cell calculation.
\label{cell}}
\end{figure}

\begin{figure}
\caption{The averaged largest, middle, and smallest components
$u_1$, $u_2$, $u_3$ and amplitude $u$ of local modes as a function of
temperature in the
cooling-down simulation of a $12\times 12\times 12$ lattice described
in Section IV.
The dotted lines are guides to the eye. \label{u-T}}
\end{figure}

\begin{figure}
\caption{The averaged homogeneous strain $\eta_H$
as a function of temperature in the
cooling-down simulation of a $12\times 12\times 12$ lattice described
in Section IV.
The strains are measured relative to the LDA minimum-energy cubic structure
with lattice constant 7.46 au.
The dotted lines are guides to the eye. \label{e-T}}
\end{figure}

\begin{figure}
\caption{The calculated pressure-temperature phase diagram.
The cubic-tetragonal (C-T), tetragonal-orthorhombic (T-O)
and orthorhombic-rhombohedral (O-R) transitions are labeled by solid
triangles, circles and squares, respectively.
The vertical dash-dot line at $P$=$-$4.8 GPa corresponds to zero pressure in
the experiment to compensate for the LDA volume error.
\label{p-T}}
\end{figure}

\begin{figure}
\caption{The probability distribution of the Cartesian component of
the local mode variable $u_x$
in the cubic phase at $T=320$K (solid line), $350$K (dashed line), and
$500$K (dotted line).\label{udos}}
\end{figure}

\begin{figure}
\caption{Temperature dependence of squared eigenfrequency
$\omega^2$ at ${\bf k}=(\pi/4a,\pi/4a,0)$ for
(a) LO, (b) TO1, and (c) TO2 modes.
\label{u2g}}
\end{figure}

\begin{figure}
\caption{Calculated mode hardness quantity
$\rho({\bf k})$, Eq.\ (\ref{rhodef}),
along special directions in the Brillouin zone.
\label{u2g2}}
\end{figure}


\begin{references}
\bibitem{latent}
Under a {\it physical} applied pressure, the latent heat would
be the change of enthalpy, not energy.  If this were applied to
our case of a {\it fictitious} pressure, the calculated latent
heats would be approximately twice as large.
\end{references}
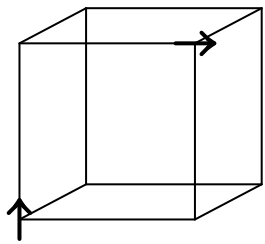
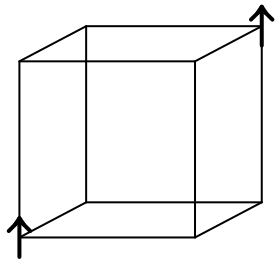
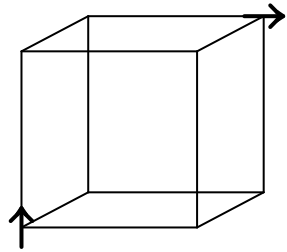
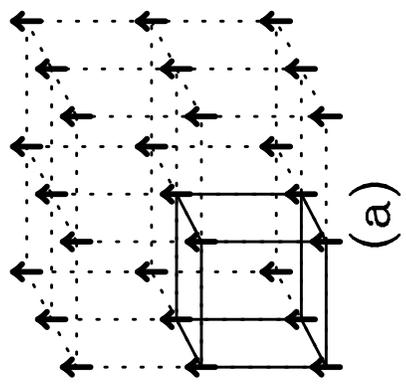
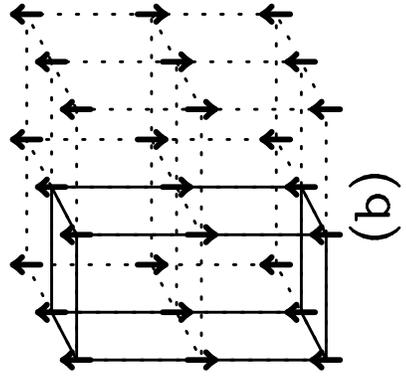
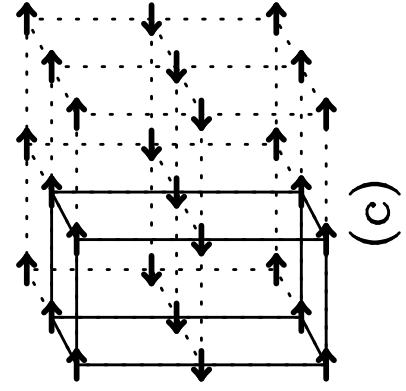
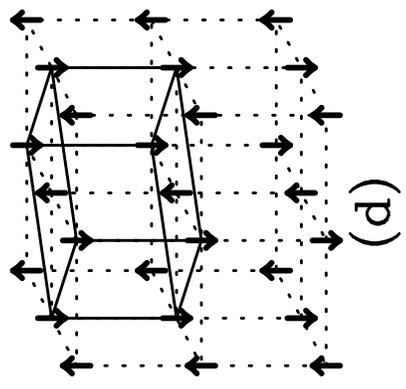
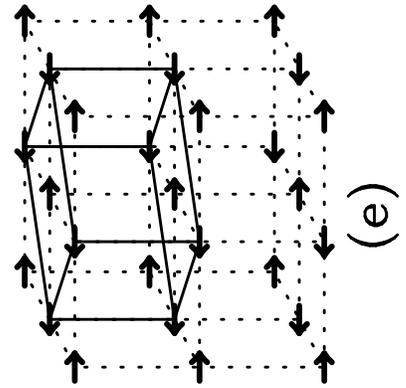
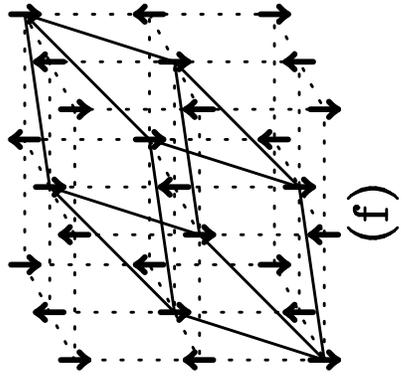
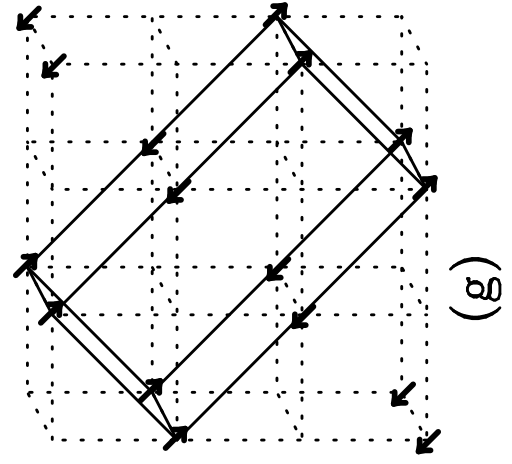
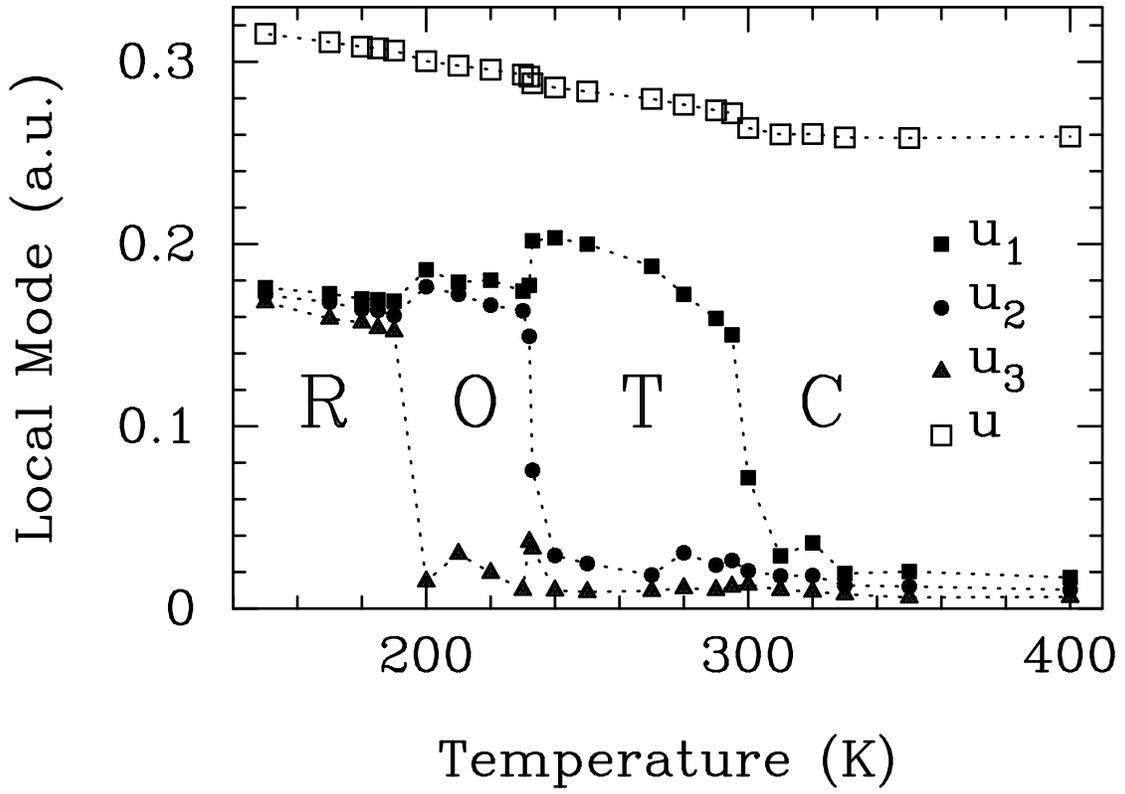
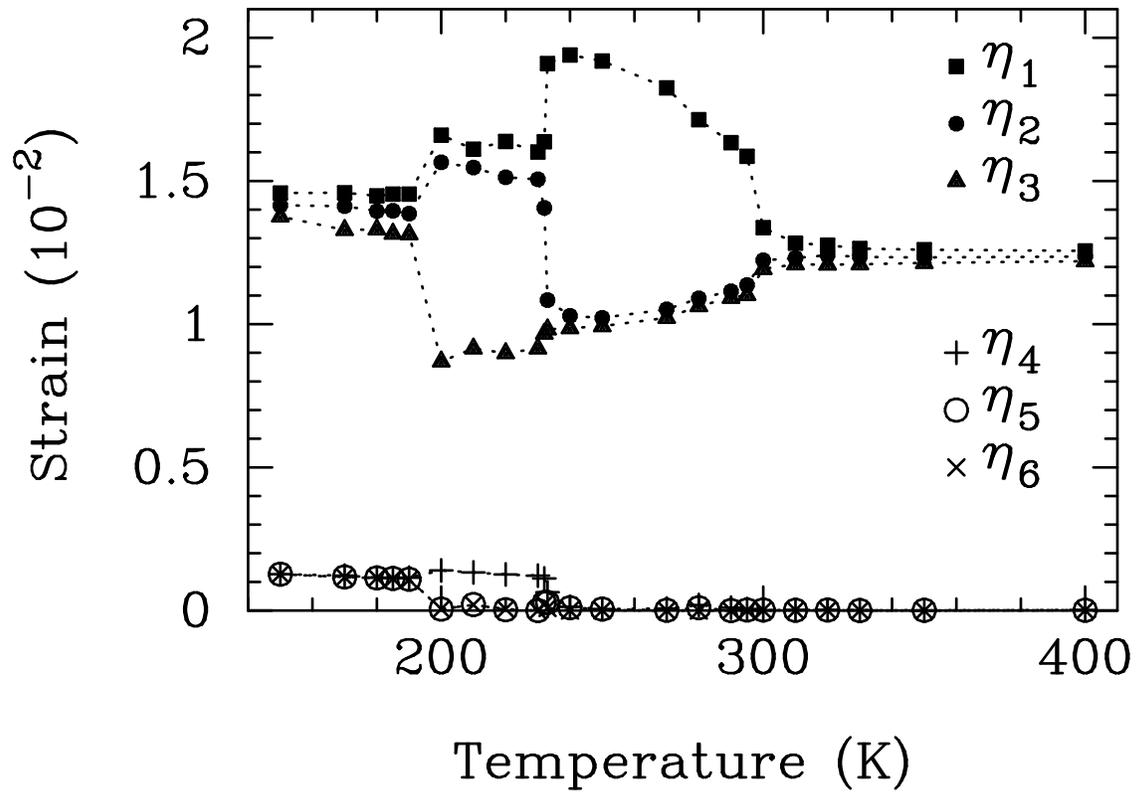
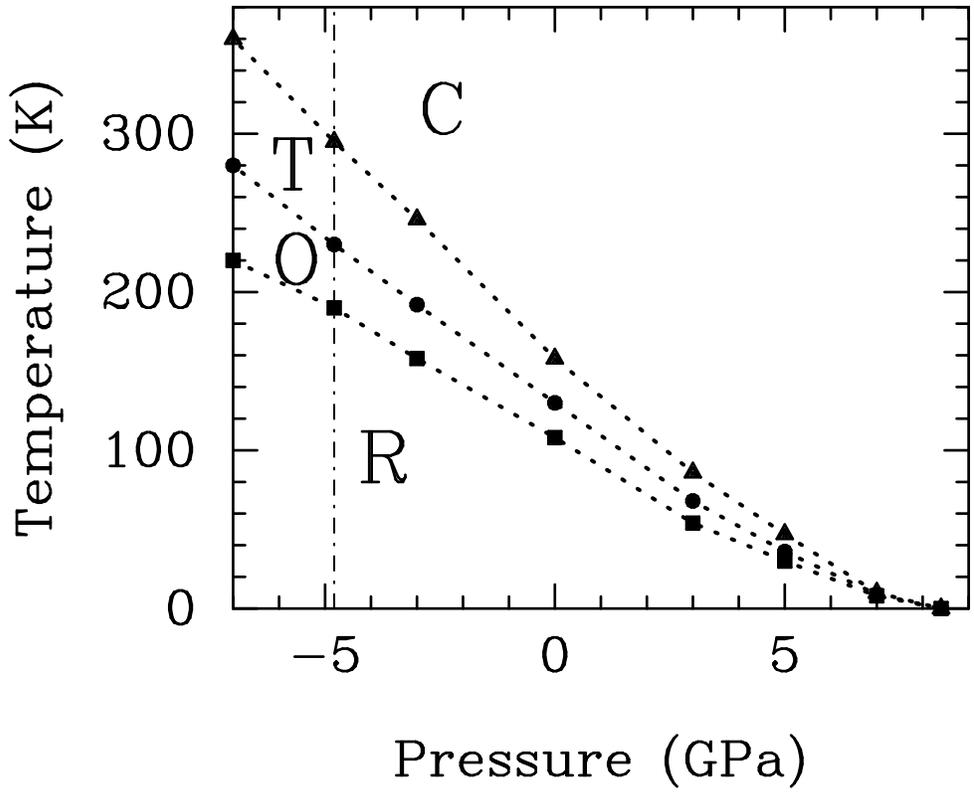
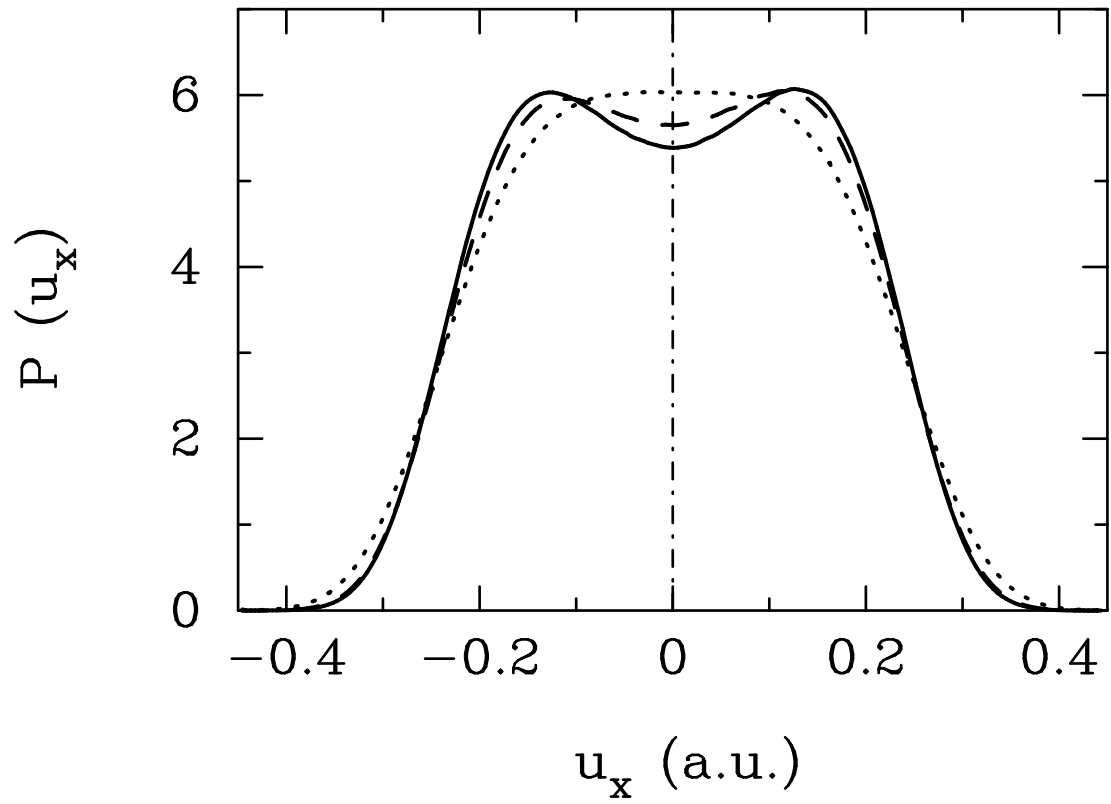
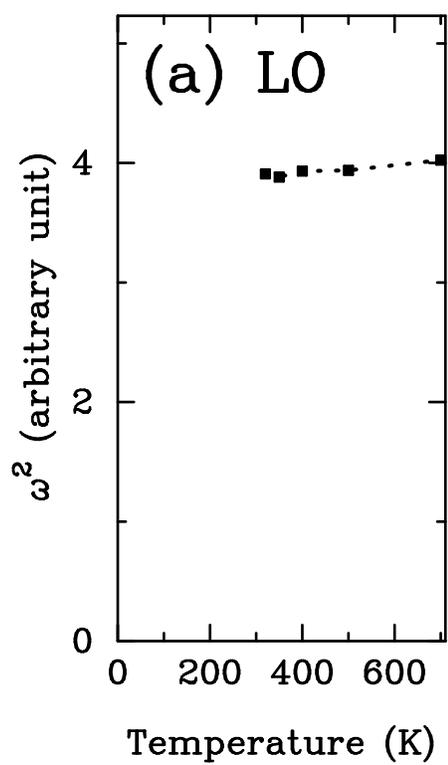
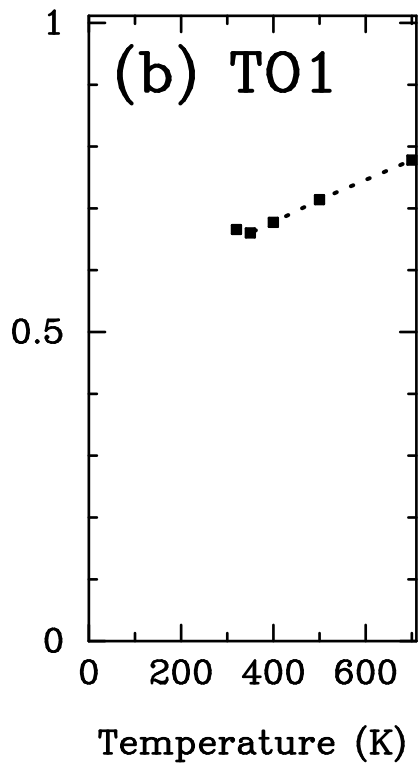
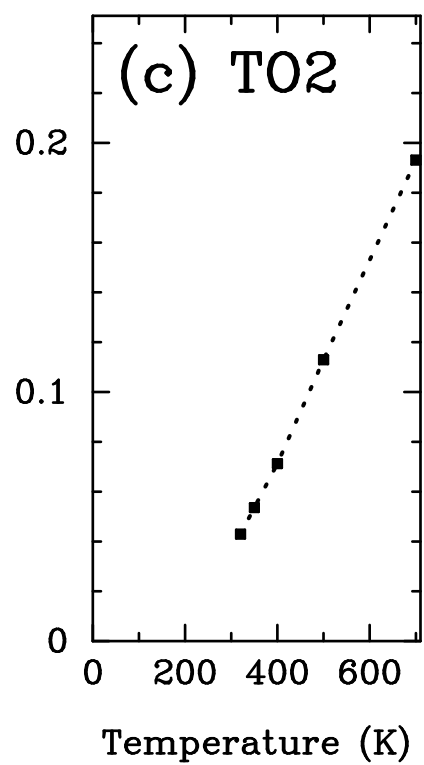
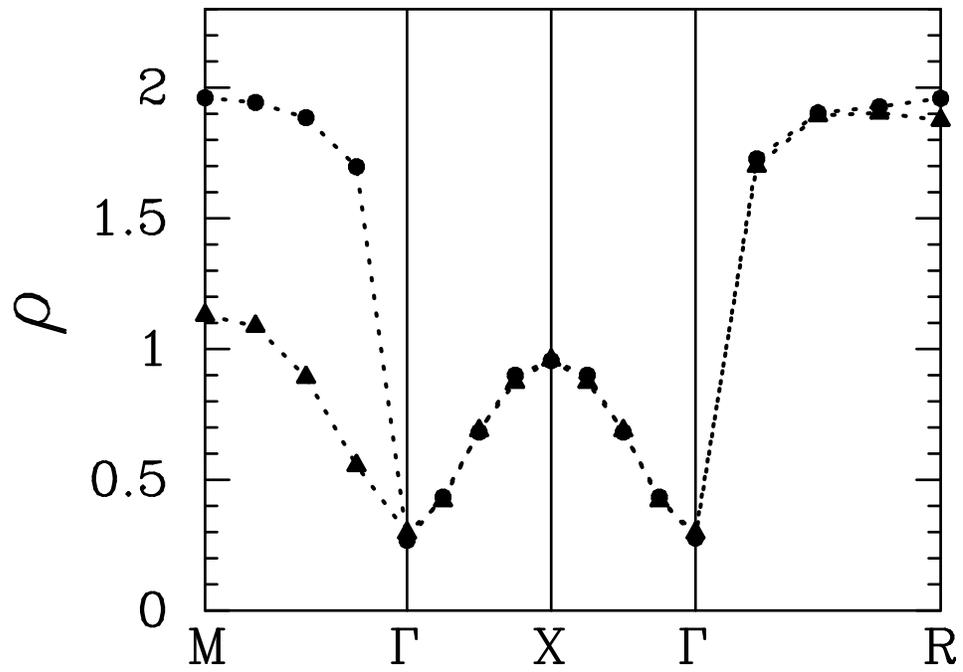
\end{document}